# V viewpoints

DOI:10.1145/3584710  Marshall Van Alstyne, Michael D. Smith, and Herbert Lin

## Economic and Business Dimensions
# Improving Section 230, Preserving Democracy, and Protecting Free Speech

*Proposing a framework for a decentralized market where no one party controls the flow of information.*

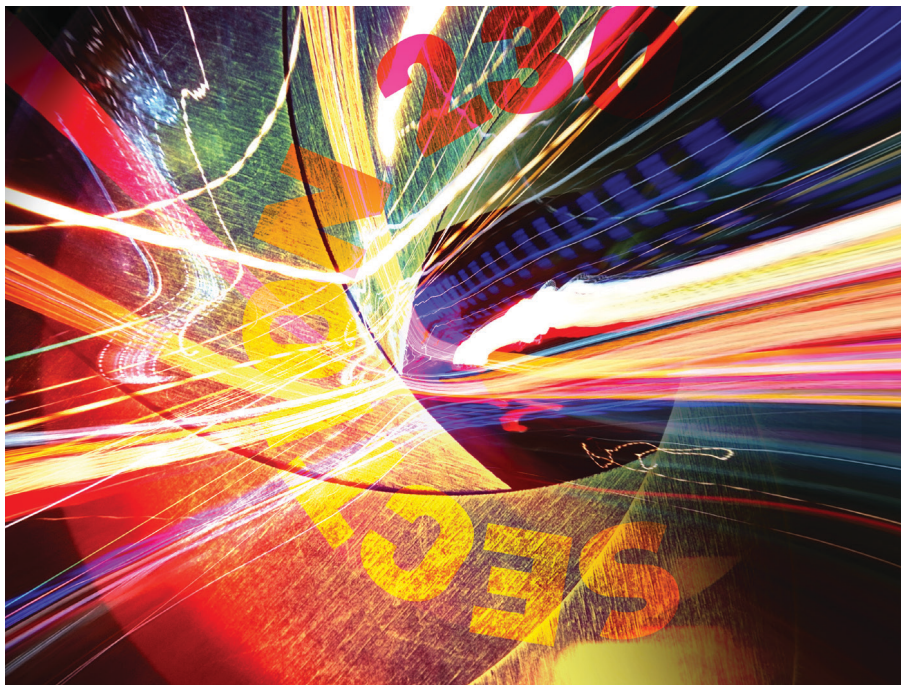

Misinformation and disinformation spread on social media have been implicated in lynching, vaccine hesitancy, polarization, insurrection, genocide, sex trafficking, drug trafficking, teenage depression, cancer misinformation, and the belief that a sitting president stole an election.[a]

The prevalence of misinformation and disinformation in online social media challenges the traditional U.S. view that in a market of many ideas, the best ones win: "The remedy for speech that is false is speech that is true."[b]

But the principle of remedying false speech with speech that is true does not work if listeners only hear unbalanced or false statements, a condition facilitated by a social media environment that algorithmically insulates listeners from hearing competing messages. The ability to remedy false speech with true speech is also hampered by the perverse incentives created by Section 230 of the Communications Decency Act. Section 230—also denoted as § 230—was enacted to provide liability protections for new market entrants at a time when their power to damage society was limited. It still does so today, but it also fosters moral hazard. Freed of consequences of their choices, social media companies worth tens of billions dollars in market capitalization act, as one whistleblower testified,[c] to benefit themselves at the expense of society.[d]

The core problem lies at the intersection of business model and law. We want social media platforms to have sufficient financial incentives to provide innovative new services. But we

---

a Strikingly, the prior sitting president was deplatformed without due process after being exempted for years from platform rules against making threats and spreading lies.
b See https://bit.ly/3I2NwTs

c See https://wapo.st/3lAjUVQ
d See https://bit.ly/3xqtFIM
<small>IMAGE BY ANDRIJ BORYS ASSOCIATES, USING SHUTTERSTOCK</small>

also want them to have sufficient legal incentives to serve the public good. These desires are often in tension.

To help resolve these tensions, we propose that changes to Section 230 should be motivated and guided by four key principles.

**First principle.** First Amendment jurisprudence should continue to reign, but within an environment that: keeps the government out of the business of defining unacceptable speech; acknowledges that intentional lies do not have the same social value as other forms of speech ("there is no constitutional value in false statements of fact. Neither the intentional lie nor the careless error materially advances society's interest in 'uninhibited, robust, and wide-open' debate on public issues,'" *Gertz v. Welch*, 1974); gives original speech a higher degree of protection than algorithmically amplified speech (for example, Aspen Commission on Information Disorder, 2021[e]); and provides enhanced content-neutral measures to facilitate counterspeech as a remedy to false speech.

**Second principle.** Users should have control (though not necessarily absolute control) over the rationale for the content they see. Today, users are subject to the rationales of the firms providing social media services. Given the network effects built on social networks, these concentrated markets offer little choice at all. Instead, users should be granted "in situ"[f] data rights, an ability to import algorithmic rationales of their choice into the infrastructure where their data resides.[g]

**Third principle.** Platforms should have some responsibility for facilitating a degree of diversity in the information environment to which individuals are exposed so that "'uninhibited, robust, and wide-open' debate on public issues" can realistically occur.

**Fourth principle.** Parties responsible for content moderation should incur some liability when content inconsistent with their moderation principles causes harm. The extent of liability should depend on the magnitude of harm caused—harmful disinformation propagated widely should

---

e See https://bit.ly/3YRHXxN
f See https://bit.ly/3xomXmz
g See https://bit.ly/3mpbQrb

---

> **The core problem lies at the intersection of business model and law.**

---

result in greater liability than the same disinformation propagated minimally. Note that since content moderation entails decisions about what content to deprecate and/or to recommend, amplification is conceptually a part of content moderation.

With these principles in mind, one possible implementation would involve changing the way users, content creators, new third-party content moderators, and social media platforms interact.

**Users** would have the freedom to choose content moderation policies they endorse. That is, user content streams would include mostly (but not exclusively) content conforming to the moderation policies they chose for themselves. One person could choose content moderation handled by BBC, another by Fox News, and another by *Consumer Reports*. With in situ data rights, users could import the content filter of their choice within a market for content moderation results.

**Content creators** would have the ability to bypass moderation policies by "warranting" their content. A warrant is an enforceable attestation that the content in question is not per se illegal (for example, does not violate copyright laws, is not child pornography, is not an immediate incitement to violence or some illegal act) and that it is not materially false (for example, the Pope endorsed my candidate, vaccines contain microchips, elections are on Wednesday not Tuesday). The content creator warrants their content by placing some specific asset at risk, an asset that is forfeited if the attestation proves to be invalid or untrue. In return, the content creator gains the right for his or her content to be privileged in delivery by the platform. This ensures "counterspeech" can pierce any filter bubble a content moderation policy might create. A party objecting to the specifics of the warranted content is free to challenge the warrant, and if the content creator loses the challenge, the challenger wins the warrant. These attestations have several important properties: Liars do not want to warrant false facts; fake news becomes costlier than honest journalism; and truth tellers can pierce filter bubbles.[h]

**Third-party content moderators** would be private parties responsible for shaping the content feeds of users based on clearly stated and algorithmically enforced standards that define unacceptable content.[i] The goal is a decentralized market of moderators each competing to offer better filters. Advertisers would contract with content moderators rather than platforms, based on their particular moderation policies, to avoid association with content they did not like. Moderators would be free to make money based on ads or subscription as they like.

Realistically, it is not possible to guarantee no unacceptable content ever reaches users, but moderation standards could include a maximum acceptable error rate above which the moderator becomes liable for negligence. At their discretion, content moderators could block algorithmically amplified content unless it were warranted as described in this column. Warranted content shifts liability to the content creator: Creators do not need to warrant content, and moderators do not need to pass non-warranted content to their users.

**Social media platforms** would continue to provide distribution services, but would be required to change their practices in certain ways. First, they must offer users an unmoderated option, which would provide the user only with content from sources selected explicitly by the user. Second, platforms must offer Application Programming Interfaces to third-party moderating software that allow them to implement the services described in this column. Third, platforms must enter into contracts with moderators on fair, reasonable, and non-discriminatory (FRAND) terms[j] that would pay the platforms from the income that moderators gen-

---

h See https://bit.ly/3YXF6Ud
i See https://bit.ly/3xnSt48
j See https://bit.ly/3k191ff

erate from taking ads. At its own discretion, a platform could offer users its own content moderation service, but no user would be forced to use it.

This arrangement would significantly alter the perverse incentives present today. Advertisers might need to engage with multiple content moderators rather than a single platform but gain greater brand identity through moderator affiliation. Users would have the freedom to decide which content moderation platform(s) to use. Creators could reach any audience with any content they are willing to certify. Platforms would be free to offer their own content moderation policies, but users would make the final determination about which policy to use. Dispersing choices reduces the central role that platforms play today, a feature not a bug, when our goal is to decentralize content moderation decisions.

Our proposed implementation depends critically on the ability to hold content moderators accountable for their actions (or their failure to act). Such accountability requires human judgments about when content improperly passed or blocked through a filter, and Van Alstyne has described a decentralized system of juries to make such judgments.[k]

Proposals for accountability have faced two other critiques. The first critique is that holding content moderators accountable for false speech will have the inevitable side effect of chilling legitimate speech. However, the separation of original speech from algorithmically amplified speech enables content creators to post freely.

The second critique is that the volume of content, especially when amplified algorithmically, is too large to make judgments about any individual instance. However, content moderators can be held accountable on a statistical basis rather than on its performance in individual instances. The Central Limit Theorem (CLT)[l] guarantees that establishing the presence of misinformation in amplified speech is feasible to any level of desired accuracy simply by taking larger samples. A progressive error rate for material improperly passed or blocked by a moderator's filter easily adjusts to big firms and small. If Facebook's content moderation is allowed 1% error rate, perhaps startup content moderators are allowed 5%. Facebook, for example, already reports removal statistics for suicide and self-harm,[m] COVID misinformation,[n] terrorist content,[o] and regulated goods.[p]

> **Realistically, it is not possible to guarantee no unacceptable content ever reaches users.**

How should liability be allocated for failure to abide by a moderator's performance standards? The pollution problem is one of joint choices: one party posts and another amplifies. The status quo solution is via lawsuits to determine allocation, but those are unpredictable, expensive, and slow. However, separating original and amplified content circumvents the allocation of liability. Charging moderators more than the ad revenues they generate from amplifying untrue and unwarranted content makes doing so unprofitable, even as the original content remains. The social cost of the unamplified original content is not eliminated but can be regarded as the cost of preserving free speech rights.

We acknowledge our proposal addresses only a fraction of the entire problem set posed by pollution in the information environment. A common understanding is that such pollution is primarily a matter of truth. Misinformation is information that is false, and disinformation is misinformation disseminated with the knowledge that it is false. Yet much of the damage is not truth or falsity but externalities in the form of information-driven lynchings, insurrections, and sex trafficking that occur off platform. These are the information pollution today, like the industrial pollution a century ago, that need proper liability assessment.

Parody and "lawful but awful"[q] speech also go far beyond claims that are true or false. Fiction intentionally depicts events that never happened. Exhortations, opinions, and questions are neither true nor false—as examples, consider statements such as "Kill all the [insert group name here]," or "Republicans are more patriotic than Democrats," or "Didn't I see you at a Nazi rally?" A statement can have entirely different implications depending on which words are emphasized—try different emphases for the words in *"I never said you stole that money."* Speakers can claim that their comments were intended to be humor or sarcasm rather than to be taken seriously. Conspiracy theories cannot be falsified when it is impossible to prove the negative.

Our goal in this proposed framework has been to create a decentralized market where no one party—not government, not private enterprise, not powerful individuals—control the flow of information.[r] The principles outlined in this column of honoring the First Amendment, giving users control, giving content creators control, and balancing liability assignment seek to do just that. Separating original post from amplification helps identify where to place liability. Enabling attestation creates a credible signal of whether an author is willing to accept accountability. Our hope is that changes consistent with these principles will enable the creation of a more decentralized and democratic instrument that is fairer and more predictable and operates at lower cost to limit societal damage and avoid lawsuits.

---

k  See https://bit.ly/3k10ZD5
l  See https://bit.ly/3L45ZlN
m  See https://bit.ly/3XCq3Oo
n  See https://bit.ly/3XCqW9G
o  See https://bit.ly/3KLFhOk
p  See https://bit.ly/3kA8G3p
q  See https://bit.ly/3YzEDYh
r  See https://bit.ly/3k10ZD5

---

**Marshall Van Alstyne** (mva@bu.edu) is a Questrom Chair Professor at Boston University, Boston, USA, where he teaches information economics. He is also a Digital Fellow at the MIT Initiative on the Digital Economy and co-author of the international best-seller *Platform Revolution* (W.W. Norton).

**Michael D. Smith** (mds@cmu.edu) is the J. Erik Jonsson Professor of Information Technology and Marketing at Carnegie Mellon University in Pittsburgh, PA, USA.

**Herbert Lin** (herblin@stanford.edu) is a senior research scholar and Hank J. Holland Fellow at Stanford University, Stanford CA, USA, where he focuses on emerging technologies and national security.